\documentclass[a4paper,12pt]{iopart}

\usepackage{graphicx}
\usepackage{amssymb}
\usepackage{epstopdf}

\begin{document}

\title[New transfer-matrix algorithm for exact enumerations]{A new transfer-matrix algorithm for exact enumerations: 
Self-avoiding polygons on the square lattice}
\author{Nathan Clisby and Iwan Jensen}
\address{\small ARC Centre of Excellence for Mathematics and Statistics of Complex Systems,
Department of Mathematics and Statistics,
The University of Melbourne, VIC 3010, Australia}
\begin{abstract}
We present a new and more efficient implementation of transfer-matrix
methods for exact enumerations of lattice objects. The new method is illustrated 
by an application to the enumeration of self-avoiding polygons on the square lattice. 
A detailed comparison with the previous best algorithm shows significant 
improvement in the running time of the algorithm. The new algorithm is used
to extend the enumeration of polygons to length 130 from the previous record of 110.
\end{abstract}


\section{Introduction}
\label{sec:intro}
 
An {\em $n$-step self-avoiding polygon} (SAP) ${\bf \omega}$ on a regular lattice is 
a sequence of {\em distinct} vertices $\omega_0, \omega_1,\ldots , \omega_n$ 
such that each vertex is a nearest neighbour of its predecessor and 
 $\omega_0$ and $\omega_n$ are nearest-neighbours. SAP are
considered distinct up to translations of the starting point $\omega_0$ and orientation.
The SAP problem on regular lattices is one of the most important 
and classic combinatorial problems in statistical mechanics \cite{Madras93,Hughes95,PolygonBook}. 
SAP are often considered in the context of lattice models of ring-polymers and vesicles \cite{Leibler87,Fisher89,Fisher91}. 
The fundamental problem is the calculation  of the number 
of SAP, $p_n$, with $n$ steps. Note that on the square
lattice polygons have an {\em even} perimeter and $p_n =0$ for $n$ odd. 
 As most interesting combinatorial problems, SAP have exponential growth
\begin{equation}\label{eq:coefgrowth}
p_n = B\mu^n n^{\alpha-3}[1+o(1)], 
\end{equation}
\noindent
where $\mu$ is the so-called connective constant, $\alpha = 1/2$ is a (known) universal 
critical exponent \cite{Nienhuis82,Nienhuis84}, and $B$ is a critical amplitude. 
When analysing the data it is often convenient to use the associated generating function
\begin{equation}\label{eq:genfunc}
 P(x) = \sum_n p_n x^n = \widehat{B}(x)(1-x\mu)^{2-\alpha}.
\end{equation}
This series has a singularity at the critical
point $x_c = 1/\mu$ with critical exponent $2-\alpha$. 

The enumeration of SAP has a long and glorious history. Suffice to 
say that early calculations were based on various direct counting algorithms
of exponential complexity, with computing time $T(n)$ growing asymptotically 
as $\lambda^n$, where $\lambda = \mu \sim 2.638$, the connective constant 
for SAP on the square lattice. Enting \cite{Enting80} was the first to produce a major breakthrough 
by applying transfer matrix (TM) methods to the enumeration of SAP on finite 
lattices. This so-called finite lattice method (FLM) led to a very significant 
reduction in complexity to $3^{n/4}$, so $\lambda = \sqrt[4]{3}=1.316\ldots$. 
More recently we \cite{Jensen99} refined the algorithm using the method of pruning 
and reduced the complexity to $1.2^n$.

\begin{figure}
\begin{center}
\includegraphics[scale=0.8]{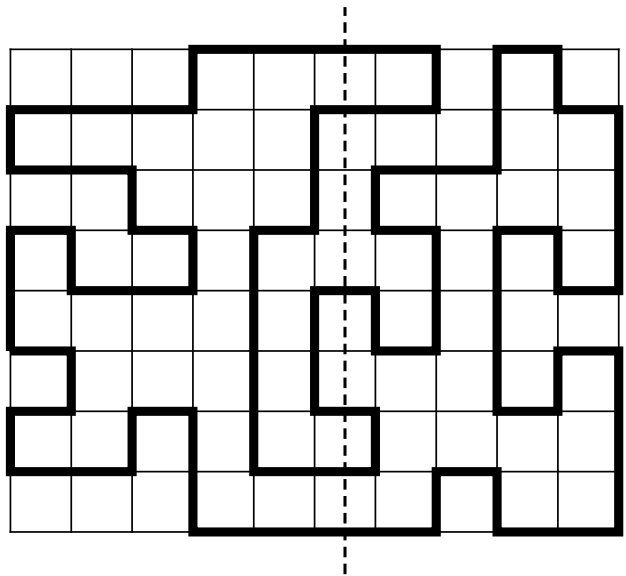}
\caption{\label{fig:sap} 
An example of a SAP on a $10\times8$ rectangular piece of the square lattice.
The dashed line shows a position of the TM boundary at a particular stage of the calculation.}
\end{center}
\end{figure}

All of the above TM algorithms are based on keeping track of the way
partially constructed SAP are connected to the left of a line bisecting
the given finite lattice (rectangles in the case of the square lattice).
In this paper we take a new approach and instead keep track of how
partially constructed SAP must connect up to the right of the boundary
line. The major gain is that it is now straightforward to calculate the
number additional bonds required to complete a given partial SAP, this
in turn results in a substantially faster algorithm. The draw-back is
that some updating rules become much more complicated. The basic idea
can best be illustrated by considering the specific example of a SAP
given in Figure~\ref{fig:sap}. If we cut the SAP by a vertical line
(the dashed line) we see that the SAP is broken into two pieces to the
left and right of the cut-line. On {\em either} side of the line we are
left with a set of partial loops. This means that at any stage a given
configuration of occupied edges along the cut-line can be described in
two ways. We can describe how the edges are connected in pairs forming
loops to the left or right of the cut-line. Moving from left to right we
can in other words keep track of what happened in the past, that is how
loops are connected to the left, or prescribe what must happen in the
future, that is how edges are to be connected to the right of the
cut-line so as to form a valid SAP.

One may think of the new FLM method as taking an initial seed SAP which
must touch the left hand boundary of the enclosing rectangle. Here there
is a known part consisting of two edges to the left of the boundary
line, and to the right of the boundary line a self-avoiding walk of
known topology but unknown shape and length. As the boundary line is
moved through the lattice the overall topology of the SAP must be
preserved, but the topology of the configuration on the right hand side
of the boundary line may be deformed.  The restriction in the extent to
which the topology can be deformed is a consequence of the choice in
updates for the boundary line, which for the square lattice only moves
to enclose one additional vertex and two additional edges in each
update.

This picture of the new FLM method is equally applicable to the
enumeration of any lattice object with fixed topology, such as
self-avoiding walks, theta graphs \cite{Guttmann78}, or star polymers.

\section{The finite-lattice method and TM algorithms}

All TM algorithms used to enumerate SAP on the square lattice build on the 
pioneering work of Enting \cite{Enting80} who enumerated square lattice 
self-avoiding polygons using the finite lattice method. 
The first terms in the series for the polygon generating 
function can be calculated using transfer matrix techniques to count 
the number of polygons in rectangles $W$ vertices wide and $L$ vertices long. 
Due to the symmetry of the square lattice one need only consider rectangles 
with $L \geq W$. Any polygon spanning such a rectangle 
has a perimeter of length at least $2(W+L)-4$. By adding the contributions 
from all rectangles of width $W \leq W_{\rm max}$  (where the choice of 
$W_{\rm max}$ depends on available computational resources) and length 
$W \leq L \leq 2W_{\rm max}-W+1$, with contributions from 
rectangles with $L>W$ counted twice, the number of polygons per vertex of an 
infinite lattice is obtained correctly up to perimeter $N=4W_{\rm max}-2$.

\subsection{Outline of the traditional TM algorithm}

\begin{figure}[ht]
\begin{center}
\includegraphics[scale=0.7]{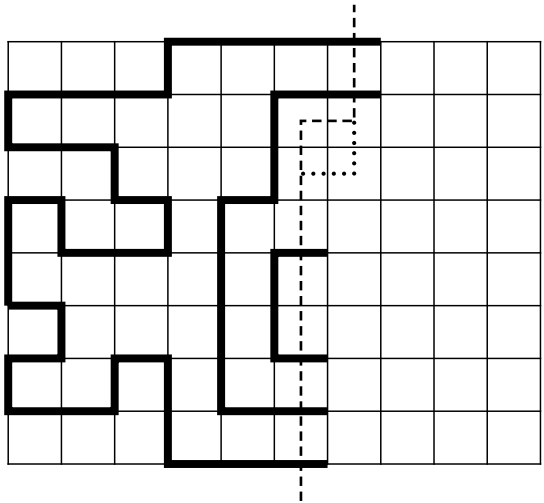}
\end{center}
\caption{\label{fig:transfer}
A snapshot of the boundary line (dashed line) during the transfer matrix
calculation of type $A$ configurations on a strip of size $8\times 10$.
SAP are enumerated by successive moves of the kink in the boundary line,
as exemplified by the position given by the dotted line, so that one
vertex and two edges at a time are added to the strip. To the left of
the boundary line we have drawn an example of a partially completed
SAP.}
\end{figure}

The generating function for a rectangle is calculated using transfer
matrix techniques. Details of our previous algorithm can be found
in \cite{Jensen99,Jensen03} and Chapter 7 of \cite{PolygonBook}. Here
we outline those aspects common to both algorithms that are needed to
appreciate the differences. The most efficient implementation of the TM
algorithm generally involves bisecting the finite lattice with a
boundary and moving the boundary in such a way as to build up the
lattice vertex by vertex as illustrated in Figure~\ref{fig:transfer}.
If we draw a SAP and then cut it by a line we observe that the partial
SAP to the left of this line consists of a number of loops connecting
two edges (we shall refer to these as loop-ends) in the intersection.
Each end of a loop is assigned one of two labels depending on whether
it is the lower or upper end of a loop. Each configuration along the
boundary line can thus be represented by a set of edge states
$\{\sigma_i\}$, where

\begin{equation}\label{eq:states}
\sigma_i = \left\{ \begin{array}{rl}
0 &\;\;\; \mbox{empty edge}, \\ 
1 &\;\;\; \mbox{lower loop-end}, \\
2 &\;\;\; \mbox{upper loop-end}. \\
\end{array} \right.
\end{equation}
\noindent
If we read from the bottom to the top, the configuration or signature $S$ along the 
intersection of the partial SAP in Figure~\ref{fig:transfer} is $S=\{1110200022\}$. 
Since crossings are not permitted this encoding uniquely describes 
how the loop-ends are connected.

In applying the transfer matrix technique to the enumeration of polygons
we regard them as sets of edges on the finite lattice with the properties:
\begin{itemize}
\item[(1)] A weight $x$ is associated with each occupied edge.
\item[(2)] All vertices are of degree 0 or 2.
\item[(3)] Apart from isolated sites, the graph has a single connected
component.
\item[(4)] Each graph must span the
rectangle from left to right and from bottom to top.
\end{itemize}

Constraint (1) is trivial to satisfy. The sum over all contributing
graphs (valid SAP) is calculated as the boundary is moved through the
lattice. For each configuration of occupied or empty edges along the
intersection we maintain a generating function $G_S$ for partial
polygons with signature $S$. In exact enumeration studies $G_S$ is
a truncated polynomial $G_S(x)$ where $x$ is conjugate to the
number of occupied edges. In a TM update each source signature $S$
(before the boundary is moved) gives rise to one or two new target
signatures $S'$ (after the move of the boundary line) and $k=0, 1$ or 2
new edges are inserted leading to the update
$G_{S'}(x)=G_{S'}(x)+x^kG_S(x)$.

Constraint (2) is easy to satisfy. If both kink edges are empty we can
leave both new edges empty or insert a partial new loop by occupying 
both of the new edges. If one of the kink edges is occupied then one
of the new edges must also be occupied. If both of the kink edges are occupied
both of the new edges must be empty. It is easy to see that these rules
leads to graphs satisfying constraint (2).

\begin{figure}[ht]
\begin{center}
\includegraphics[scale=1.0]{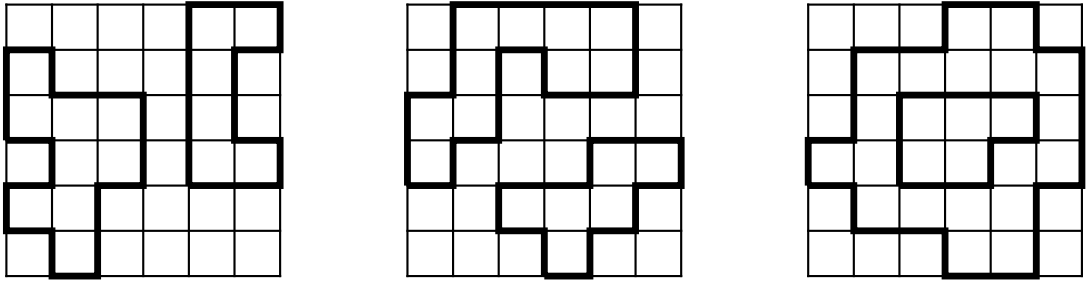}
\end{center}
\caption{\label{fig:mulcomp}
Three ways in which graphs with separate components could occur.}
\end{figure}

Constraint (3) is the most difficult to satisfy. We have shown some
examples of two-component graphs in Figure~\ref{fig:mulcomp}. Graphs of
the type shown in the left-most panel, where separate components occur
side by side, are quite easy to avoid by never allowing the insertion of
a new loop into the totally empty configuration except for the initial
seed state. This also ensures that
all polygons touch the left-most border of the rectangle. There are only
two distinct ways in which a pair of loops can be placed relative to one
another -- side by side or nested -- as shown in the last two panels of
Figure~\ref{fig:mulcomp}. With the loop encoding given above multiple
components are avoided by not allowing two connected loop-ends to join
{\em except} when no other loop-ends are present in which case a
completed SAP is formed. 

In order to satisfy constraint (4) we need to add more information to a
signature. In addition to the usual labelling of the intersection with
the boundary line we also have to indicate whether the partially
completed polygon has reached neither, both, the lower, or the upper
boundaries of the rectangle. In order to represent a given partial
polygon we have to add a some information to the usual set of edge
states $\{\sigma_i \}$. We add two extra `virtual' edge states
$\sigma_b$ and $\sigma_t$, where $\sigma_b$ is 0 or 1 if the bottom of
the rectangle has or has not been touched. Similarly, $\sigma_t$ is 0 or
1 if the top of rectangle has or has not been touched.

\subsection{The new algorithm}

Most of the basic properties and considerations outlined in the previous
Section apply also to the new algorithm. A major difference is
conceptual namely that as stated in Section~\ref{sec:intro} we change
the way we keep track of the partial loops intersecting the boundary
line. While we can use exactly the same encoding of a signature as
(\ref{eq:states}) the meaning of `lower' and `upper' loop-end is
profoundly different. In the original algorithm these terms referred to
partial loops connected to the left of the boundary, that is, to how
the existing loops in the partially generated SAP are already
connected. However, in the new algorithm `lower' and `upper' loop-end
refers to how occupied edges along the boundary must be connected via a
loop at some later stage (to the right of the boundary). This change in
turn results in new updating rules for the cases where a new loop is
inserted or two loop-ends join at the kink. We deal with the latter
easier modification first.

We can join two loop-ends at the boundary kink {\em only} if they belong
to the same loop, thus closing a partial loop of the SAP, since the loop
encoding of new algorithm simply prescribes how occupied edges are to
be connected. Thus the only valid case is the kink-state `12' and all
the other kink states (`11', `22' and `21') are forbidden since they
would correspond to connecting occupied edges which should not have been
connected. Two situations arise when a partial loop is closed;
{\em either} there are other occupied edges along the boundary and one
just proceeds with the calculation or all other edges are empty and a
closed SAP is formed and added to the running total for the SAP
generating function.

\begin{figure}[ht]
\begin{center}
\includegraphics[scale=1.0]{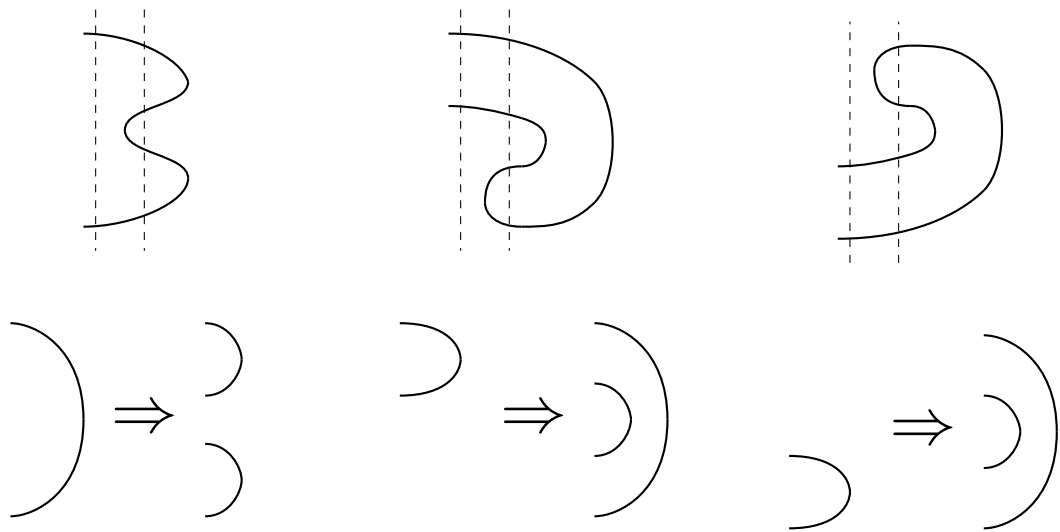}
\caption{\label{fig:basic} 
The possible basic deformations to the topology of a boundary state as the
boundary line is shifted are shown schematically above. The
corresponding basic loop updates are shown immediately below.}
\end{center}
\end{figure}

While edge-joining is simplified the insertion of a new loop becomes
much more complicated. In the original approach the insertion was done
and no further action was required. In the new approach we must connect
the two new occupied edges to other occupied edges on the boundary line.
The restrictions on SAP enumeration mean that the two new occupied
edges must connect to existing {\em connected edges} provided these are
reachable (more on this later). The state of the new occupied edges
will depend on their placement relative to the edges they become
connected to, and the state of the existing occupied edges may change.
In Figure~\ref{fig:basic} we show the two basic situations: The new
occupied edges are either placed inside the loop formed by the two
existing connected edges or they are placed outside them. In the first
case, shown to the left in Figure~\ref{fig:basic}, the upper (lower) end
of the inserted loop must connect with the upper (lower) end of the
existing loop, in terms of the edges involved the states change from
`1002' to `1212'. In the second case, in the middle of the figure, the
upper (lower) end of the inserted loop must connect with the lower
(upper) end of the existing loop, in terms of the edges involved the
states change from `1200' to `1122'. So {\em both} new occupied edges
become `lower' loop-ends while the existing lower loop-end is changed to
an upper loop-end. Shown to the right in Figure~\ref{fig:basic}, there
is also a symmetric case where the new loop is placed above the existing
loop leading to the state change `0012' to `1122'.

\begin{figure}
\begin{center}
\includegraphics[scale=0.7,angle=90]{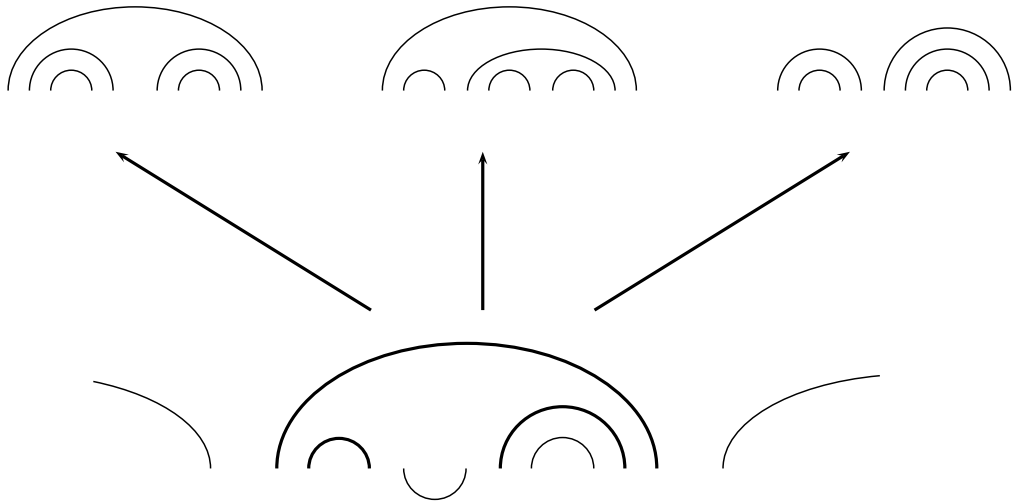}
\caption{\label{fig:insert} 
The possible updates resulting from the insertion of a new partial loop
into an existing loop configuration. At the bottom we indicate by a
lower arc the new partial loop. In the existing loop configuration
(upper arcs) accessible loops are indicated with heavy lines. The three
possible new loop configurations are shown on top.}
\end{center}
\end{figure}

The newly inserted loop can connect to any existing loop that can be
reached without crossing another loop. The general situation is
illustrated in Figure~\ref{fig:insert} where we see that the new loop
can be connected to three existing loops (indicated by thick lines).
The second loop to the right of the new loop is nested inside an
existing loop and can therefore {\em not} be reached without crossing
the enclosing loop. Likewise any loops outside the large loop enclosing
the new loop are unreachable. So in this case the insertion of a single
new loop gives rise to three new loop configurations as illustrated in
the top panels of Figure~\ref{fig:insert}. The states of the edges in
the new loop configurations are obtained by applying the appropriate
basic loop insertion from above.

At this stage it seems nothing has been gained. Some updates simplify
while loop-insertion becomes much more complicated. The true pay-off
comes when we look to pruning. In the original approach pruning can be
very complicated. With deeply nested configurations one simply has to
search through all possible ways of connecting existing partial loops in
order to find the connection pattern which minimises the number of extra
edges required to form a valid SAP. In the new approach this
complication is completely gone since connections between edges is
already prescribed: There is one and only one way of completing the SAP! 

\subsection{Pruning}

The principle behind pruning is quite simple and briefly works as follows.
Firstly, for each signature we keep track of the current minimum 
number of steps $n_{\rm cur}$ already inserted to the left of the boundary 
line in order to build up that particular configuration. Secondly, we 
calculate the minimum number of additional steps $n_{\rm add}$ required to 
produce a valid polygon. There are three contributions, namely the number 
of steps required to close the polygon, the number of steps needed (if any) 
to ensure that the polygon touches both the lower and upper border, and 
finally the number of steps needed (if any) to extend at least $W$ edges 
in the length-wise direction (remember we only need rectangles
with $L \geq W$). If the sum $n_{\rm cur}+n_{\rm add} > N=4W_{\rm max}+2$ we 
can discard the partial generating function for that configuration,
and of course the configuration itself, because it will not make a 
contribution to the polygon count up to the perimeter lengths we are 
trying to obtain. For instance polygons spanning a rectangle with a width 
close to $W_{\rm max}$ have to be almost convex, so very convoluted 
polygons are not possible. Thus configurations with 
many loop-ends (non-zero entries) make no contribution at perimeter 
length $\leq N$.

\begin{figure}
\begin{center}
\includegraphics[width=11cm]{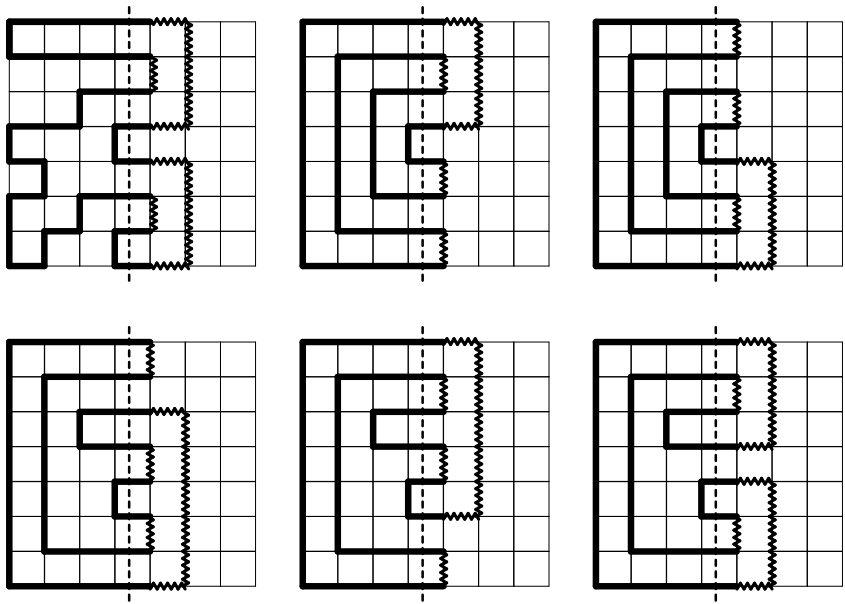}
\end{center}
\caption[Possible completions of partially generated polygons]
{\label{fig:sapclose} Examples of partially generated
polygons (thick solid lines) to the left of the intersection (dashed line)
and how to close them in a valid way (thick wavy line). Upper left panel 
shows how to close the configuration $\{12112212\}$. The upper middle and
right panels show the two possible closures of the configuration
$\{11112222\}$. The lower panels show the three possible closures of the 
configuration $\{11121222\}$. }
\end{figure}

The complicated part of the pruning approach is the algorithm to
calculate the number of steps required to close the polygon. In the
first stage we connect any separate pieces as illustrated in
Figure~\ref{fig:sapclose}. Separate pieces are easy to locate and
(provided one is not at the last edge in the configuration) the top-most
upper edge of one piece can be connected to the lower edge above. Then
$n_{\rm add} $ is incremented by the number of steps (the distance)
between the edges and the two edge-states are removed from the
configuration before further processing. In the second stage we
transform configurations starting (ending) as $\{112\ldots 2\}$
($\{1\ldots 122\}$) since the two lower (upper) edges can safely be
connected. The two edge-states are removed from the configuration --
leading to the new configuration $\{001\ldots 2\}$ ($\{1\ldots 200\}$)
-- before further processing. 

After these two stages we may be left with a configuration which has
just a single lower edge and a single upper edge. We are almost done
since these two edges can be connected to form a valid polygon. This is
illustrated in Figure~\ref{fig:sapclose} where the upper left panel
shows how to close the partial polygon with the intersection
$\{12112212\}$, which contain three separate pieces. After connecting
these pieces we are left with the configuration $\{10012002\}$. We now
connect the two lower edges and note that the first upper edge is
relabeled as a lower-edge (it has become the new lower end of the loop).
Thus we get the configuration $\{00001002\}$ and we can now connect the
remaining two edges and end up with a valid completed polygon. Note that
in the last two cases, in addition to the steps spanning the distance
between the edges, an additional two horizontal steps had to be added in
order to form a valid loop around the intervening edges. 

If the transformations above do not result in a closed polygon we must
have a configuration of the form $\{111\ldots 222\}$. The difficulty
lies in finding the way to close such configurations with the smallest
possible number of additional steps. Suffice to say that if the number
of non-zero entries is small one can easily code all possible valid ways
of closing a polygon and thus find the minimum number of additional
steps. In Figure~\ref{fig:sapclose} we show all possible ways of
closing polygons with 8 non-zero entries. Note that we have shown the
generic cases here. In actual cases there could be any number of empty
edges interspersed in the configurations and this would determine which
way of closing the SAP would minimise $n_{\rm add} $. The number
of distinct configurations are given by the Catalan numbers so there is
only 1 configuration with 6 occupied edges,  while there are 2 with 8 occupied edges,
and 5 with 10 occupied edges. In \cite{Jensen03} the various possible
ways of closing all such configurations was hand-coded. In the more
general case (configurations with 12 or more non-empty entries) we
devised a basic search algorithm that simply tried all possible ways of
closing loops. In practise initially two loop-ends are connected. The
resulting new configuration is passed through the two stages above and
we are left with a configuration with at least two fewer loop-ends. If
there are still open loops we do another pass and so on until all loops
have been closed. This process is then repeated but starting with a
different pair of initial edges. In this way one can search through all
possible ways of completing a SAP. Obviously one would expect that this
process should ultimately grow exponentially with the number of non-zero
edges. However, due to the simplifying feature of passing through the
first two stages the growth is quite slow.

In the new approach almost all of the complications of pruning are
gone. Since connections between edges are already prescribed
there is one and only one way of completing the SAP! The only
complicating factor is that in order to calculate $n_{\rm add} $ we need
to know the nesting level $l$ of each partial loop. The number of edges
it takes to connect two loop-ends at positions $i$ and $j$ is simply
$j-i+2l$. In addition we must connect to the lower and upper boundaries
and ensure that the SAP extends at least $W$ edges in the length-wise
direction. This pruning procedure can be performed in $O(W)$ operations.

\subsection{Comparative study of the algorithms}

In analysing the complexity of the two algorithms, we note that the
update step when the boundary is moved may result in $O(1)$ signatures
for the previous algorithm, and $O(W)$ for the new one. However, in the
average case we still expect the new algorithm to create $O(1)$
signatures, because connections with distant loop-ends are typically
pruned away.

For pruning, we believe that the complexity of the old algorithm is
exponential in $W$ (as we will see below, the
growth constant is mercifully small for SAP), whereas for the new
algorithm the complexity is $O(W)$.

In Table~\ref{tab:algcomp} we compare the resources used by the two
algorithms in a calculation of the number of SAP with perimeter up to
$N$. From this it is clear that the new approach is more efficient with
substantial savings in time. The required number of configurations and
terms go down very slightly while the total CPU time decrease by about
16\% for $N=98$. There is little variation in this for the listed values
of $N$. For higher values of $N$ we expect to eventually see even more
substantial savings in time. As $N$ increases more deeply nested
configurations occur and these are more expensive to prune in the old
algorithm. That we do not see an increasing time saving yet is testament
to the special nature of square lattice SAP making pruning particularly
simple for this problem, and the effort put into the original
implementation of pruning for this case. We mention in passing that a
preliminary implementation for self-avoiding walks have yielded more
substantial time savings of more than 60\% for the square lattice. 

\begin{table}[htdp]
\caption{\label{tab:algcomp} A comparison of the resources required by the
two algorithms in order to calculate the number of SAP up to length $N$.}
\begin{center}
\begin{tabular}{|r|rrr|rrr|}
\hline 
& \multicolumn{3}{c|}{Old Algorithm} & \multicolumn{3}{c|}{New Algorithm} \\
\hline
 $N$ & Configs & Terms & Time & Configs & Terms & Time \\
\hline
 42 & 3837 & 8275 & 0.31 & 3826 & 8039 & 0.23 \\
 50 & 15867 & 37389 & 2.16 & 15821 & 36660 & 1.63 \\
 58 & 61361 & 159938 & 14.60 & 59354 & 157367 & 10.75 \\
 66 & 293853 & 674548 & 95.89 & 286441 & 652805 & 78.58 \\ 
 74 & 1274667 & 3038260 & 685.41 & 1255436 & 2948937 & 571.24 \\
 82 & 4973585 & 12976379 & 4447.89 & 4921788 & 12699279 & 3783.83 \\
 90 & 22041519 & 56510740 & 26770.16 &21427764 & 54845786 & 22360.05\\ 
 98 & 94861519 & 251027714 & 173408.04 & 93020737 &244009381 & 145454.72 \\
\hline 
\end{tabular}
\end{center}
\end{table}

For large enough $N$, we expect the exponential complexity of pruning to
make the old algorithm prohibitively slow compared to the new one.
However, in practice, for conceivably achievable values of $N$, the
relative advantage of the new algorithm over the old for the enumeration
of SAP may best be described as significant rather than dramatic.  For
problems where the growth constant for pruning is large, the new
algorithm would make a dramatic difference.  Candidate problems for
potentially dramatic improvement will be mentioned in
Section~\ref{sec:summary}.

Finally we mention that due to the simplified joining of loops at the kink
(only the `12' case is permitted) it is possible to further improve the algorithm.
Since all edges are equivalent we can simplify the kink states to only consider
the four possibilities `00', `10', `20' and `12'.  That is the vertical edge is only
occupied if is is part of a loop joining at the kink. Technically 
one can also view this as replacing the two kink edges by a vertex with four possible
states.  The overall net benefit is a drop in 
memory use by about 15\% and a further improvement in running time to a net 
overall gain of some 30\%. This change also simplifies the pruning since a number
of special cases (the vertical kink edge being occupied)  are avoided. Note however
that this change is nor always permitted, i..e., on lattices with directed edges (such
as the Manhattan or $L$ lattices) the edges are equivalent and the kink-state simplification
isn't valid.

\section{Extended SAP enumeration}

The transfer-matrix algorithm is eminently suited to parallel computations and here we used the approach 
first described in \cite{Jensen03} to extend the enumeration to SAP of perimeter 130 (that is we obtain
a further 10 non-zero terms). The bulk of the calculations for this paper were performed on the cluster 
of the NCI National Facility at ANU. The NCI peak facility is a Sun Constellation Cluster
with 1492 nodes in Sun X6275 blades, each containing two quad-core 2.93GHz Intel Nehalem CPUs 
with most nodes having 3GB of memory per core (24GB per node). It took a total
of about 25000 CPU hours to enumerate SAP up to perimeter 130. We used up to 1000
processors (or more accurately cores) and up to 2.5TB of memory. Some details of resource use
are given below.
 
The integer coefficients occurring in the series expansion become very large and the calculation was
therefore performed using modular arithmetic and the series was calculated modulo various integers $m_i$ 
and then reconstructed at the end using the Chinese remainder theorem. We used the moduli 
$m_0=2^{62}$, $m_1=2^{62}-1$ and $m_2=2^{62}-3$, which allowed us to represent $p_n$ correctly. 
The NCI cluster is a heavily used shared computing facility so our major constraint was CPU
time rather than memory. For this reason we chose to perform the calculation for all $m_i$ in the
same run. Effectively this doubles the memory requirement but only results in an increase in running
time of some 15\% (compared to a run using a single $m_i$), that is, an overall {\em decrease} in
total running time by a factor of about 2.6. 

Table~\ref{tab:series} lists the new terms obtained
in this work for the number of polygons with perimeter 112--130. The number of polygons 
of length $\leq 56$ can be found in \cite{Guttmann88} while those up to length 90 are 
listed in \cite{Jensen99} and those to length 110 in \cite{Jensen03}. The full series is 
available at {\tt www.ms.unimelb.edu.au/\~{}iwan}.

\begin{table}
\caption{\label{tab:series} The number, $p_n$, of embeddings of 
$n$-step self-avoiding polygons on the square lattice. Only non-zero terms are listed.}
\begin{center}
\begin{tabular}{ll} \hline \hline
$n$ & $p_n$ \\ \hline 
112 & 646414111975777272517734370762400697757978 \\
114 & 4304591798055577073477026735313861700713176 \\
116 & 28687064652813390269800415016181829385121162 \\
118 & 191320663411431818964849556990106874938907548 \\
120 & 1276875276296096391140817393830149918943464494 \\ 
122 & 8527773411790633004325737459634720668141188468 \\
124 & 56991966408991589554333823232058663722205631080 \\
126 & 381130017241685467740337492217602004487643160168 \\
128 & 2550382601811089051031712642200910692143744745034 \\
130 & 17076613429289025223970687974244417384681143572320 \\
\hline \hline
\end{tabular}
\end{center}
\end{table}

\subsection{Resource use}

One of the main ways of achieving a good parallel algorithm using 
data decomposition is to try to find an invariant under the
operation of the updating rules. That is we seek to find some property
about the configurations along the boundary line which
does not alter in a single iteration.
The algorithm for the enumeration of polygons is quite complicated 
since not all possible configurations occur due to pruning,
and the insertion of a new loop can change the state of 
an edge far removed. However, there still is an invariant since any edge not
directly involved in the update cannot change from being 
empty to being occupied and vice versa. That is only the kink edges 
can change their occupation status. This invariant
allows us to parallelise the algorithm in such a way
that we can do the calculation completely independently on each
processor with just two redistributions of the 
data set each time an extra column is added to the lattice. Since
the number of processors we have to use is quite large (up to 1000) we
actually did 3 redistribution per column. This increases the length of the
invariant part and thus gives us better opportunities to ensure a decent
load balance. We refer to \cite{Jensen03} for details regarding the
details of our parallelised implementation.

\begin{table}[htdp]
\caption{\label{tab:para} The resources used to calculate the number of SAP on rectangles
of width $w$. Listed from left to right are the number of processors, the total CPU time in hours,
the minimal and maximal number of configurations and series terms retained and finally the
minimal and maximal time (in seconds) used in the redistribution. The minimum and maximum
is taken across all of the processors.}
\begin{center}
\small
\begin{tabular}{rrrrrrrrr} \hline \hline
$W$ & Procs & Time & Min Conf & Max Conf & Min Term & Max Term & $t$-min & $t$-max \\ \hline
23 & 160 & 241 & 9950148 & 11503231 & 73828085 & 87381144 & 560 & 1436 \\
24 & 400 & 557 & 8890870 & 9165975 & 56728718 & 61006162 & 699 & 763 \\
25 & 592 & 1125 & 11962488 &12287159 & 64909908 & 67080631 & 964 & 1096 \\
26 & 800 & 2214 & 15923256 & 16383084 & 70084896& 72376608 & 1345 & 1575 \\
27 & 800 & 3713 &24966483 & 25822571 & 87087766 &90763851 & 2321& 2650 \\
28 & 1000 & 5559 & 26089184 &26581112 &71874712 & 73445360 & 2272 & 2730 \\
29 & 800 & 4953 & 31025188 & 32021460 & 63918962 & 66526222 & 2379 & 2760 \\
30 & 400 & 2874 & 34573073 & 35446404 & 52929622 & 54330414 & 2690 & 2926 \\
31 & 96 & 450 & 25632787 & 26124867 & 31132604& 31768765 & 1482 & 1588 \\ \hline \hline
\end{tabular}
\end{center}
\end{table}

In Table~\ref{tab:para} we have listed the main resources used by the
parallel algorithm in order to enumerate SAP up to perimeter 130. For
each width $W$ we first list the number of processors used and the total
CPU time in hours required to complete the calculation for a given
width. One of the main issues in parallel computing is that of load
balancing, that is, we wish to ensure to the greatest extent possible
that the workload is shared equally among all the processors. This
aspect is examined via the numbers in columns 4--9. At any given time
during the calculation each processor handles a subset of the total
number of configurations. For each processor we monitor the maximal
number of configurations and terms retained in the generating functions.
Note that the number of terms listed is per modulo $m_i$; so in total three
times this number is actually stored. The load balancing can be roughly
gauged by looking at the largest (Max Conf) and smallest (Min Conf)
maximal number of configurations handled by individual processors during
the execution of the program. In columns 6 and 7 are listed the largest
(Max Term) and smallest (Min Term) maximal number of terms retained in
the generating functions associated with the subset of configurations.
As can be seen the algorithm is very well balanced. Finally in columns 8
and 9 we have listed the minimal and maximal total time (in seconds)
spent by any processor in the redistribution part of the algorithm and
as can be seen this part of the algorithm takes a total of some 15\% of
the CPU time. Note that most of this time is spent preparing for the
redistribution and processing the data after it has been moved. The
actual time spent in the MPI message passing routines is less than 5\%
of total CPU time.

\section{Series analysis and results}
 
To obtain the singularity structure of the generating functions we used
the numerical method of differential approximants \cite{GuttmannDA}. We
will not describe the method here and refer the interested reader to
\cite{GuttmannDA} for details, and Chapter 8 of \cite{PolygonBook} for
an overview of the method. Since all odd terms in the series are zero
and the first non-zero term is $p_4$ we actually analysed the series
$F(x)=\sum p_{2n+4} u^n$. This function has a critical point at
$u=x_c^2$ with the same exponent as that of (\ref{eq:genfunc}). In
Table~\ref{tab:analysis} we list estimates for the critical point
$x_c^2$ and exponent $2-\alpha$.  The estimates were obtained by
averaging values obtained from second and third order differential
approximants. For each order $L$ of the inhomogeneous polynomial we
averaged over those approximants to the series which used at least the
first 55 terms of the series (that is, polygons of perimeter at least
110).  The quoted error for these estimates reflects the spread
(basically one standard deviation) among the approximants. Note that
these error bounds should {\em not} be viewed as a measure of the true
error as they cannot include possible systematic sources of error. Based
on these estimates we conclude that $x_c^2 = 0.143680629269(2)$ and
$\alpha = 0.500000015(20)$.
 
Some years ago \cite{Conway93a} it was pointed out that the polynomial
$581x^4 + 7x^2 -13$ is the only polynomial with ``small'' integer
coefficients for which the relevant zero
$x_0^2=0.1436806292698685\ldots$ is consistent with the estimate for
$x_c$.  Clearly, with almost 12 digit accuracy the conjectured value
still stands.  It should be emphasised that there is no theoretical
motivation for the conjecture. However, the agreement with the numerical
estimate is very impressive (and perhaps surprising).  In any case the
polynomial can at least serve as a useful memory aid for $x_c$.

\begin{table}[htdp]
\caption{\label{tab:analysis} Estimates for the critical point
$u_c^2$ and exponent $2-\alpha$ obtained from second and third order
differential approximants to the series for square lattice
polygon generating function. $L$ is the order of the inhomogeneous
polynomial.}
\begin{center}
\small
\begin{tabular}{lllll} \hline \hline
 $L$ & \multicolumn{2}{c}{Second order DA} & 
 \multicolumn{2}{c}{Third order DA} \\ \hline 
 & \multicolumn{1}{c}{$u_c^2$} & \multicolumn{1}{c}{$2-\alpha$} & 
 \multicolumn{1}{c}{$u_c^2$} & \multicolumn{1}{c}{$2-\alpha$} \\ \hline
0 & 0.1436806292669(38)& 1.500000027(19)& 
 0.1436806292683(18)& 1.500000023(10) \\ 
2 & 0.1436806292697(35)& 1.500000021(71)&
 0.1436806292690(12)& 1.500000021(25) \\
4 & 0.1436806292684(16)& 1.500000023(14)& 
 0.1436806292694(11)& 1.5000000155(66) \\ 
6 & 0.1436806292688(18)& 1.500000019(11)&
 0.14368062926979(89)& 1.5000000131(61) \\
8 & 0.1436806292684(18)& 1.500000018(12)&
 0.1436806292699(15)& 1.500000013(10) \\
10 & 0.1436806292686(16)& 1.500000019(12)&
 0.1436806292695(11)& 1.5000000154(68) \\
 \hline \hline
\end{tabular}
\end{center}
\end{table}

\begin{figure}
\begin{center}
\includegraphics[scale=0.95]{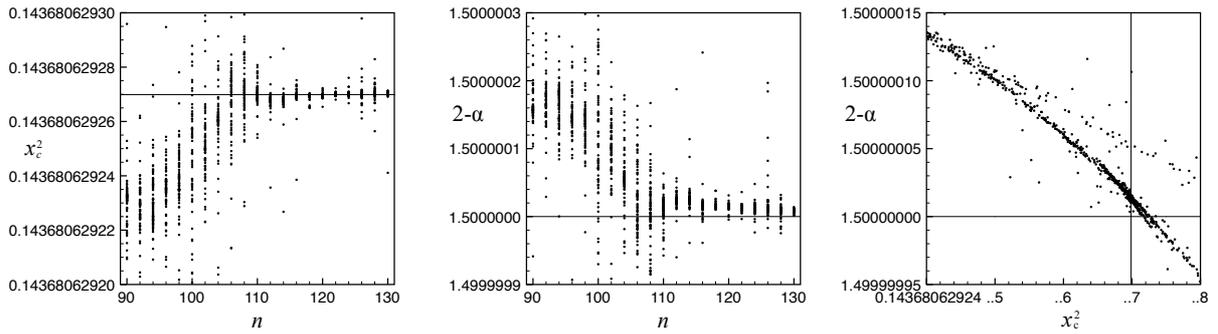}
\end{center}
\caption{\label{fig:crpexp} Estimates of the critical point $x_c^2$
and critical exponent $2-\alpha$ versus $n$ (left and middle panels)
 and $2-\alpha$ versus $x_c^2$ (right panel) for the
square lattice polygon generating function. The straight lines
correspond to $2-\alpha=3/2$ and $x_c^2=0.1436806292698685\ldots$.}
\end{figure}

 To gauge whether or not the estimates truly are as well converged as the results
 in Table~\ref{tab:analysis} would suggest we find it useful to plot the actual individual
 estimates against $n$ (where $p_n$ is the last terms used to form a given differential approximant).
 In the first two panels of Figure~\ref{fig:crpexp} we have plotted the estimates for $x_c^2$ and
 $2-\alpha$ as functions of $n$. Each point represents an estimate from
 a third order differential 
 approximant. The approximants appear very well converged and given the very high
 resolution of the abscissa there is no sign of any significant systematic drift. 
 Finally, in the third panel we plotted the estimated for $2-\alpha$
 versus the corresponding
 estimates for $x_c^2$. If the conjectures for the exact values are correct the estimates
 should ideally pass through the point of intersection between the conjectured values.
 Clearly there is a very slight discrepancy here and for $2-\alpha=3/2$ the `biased'
 estimate for the critical point is $x_c^2=0.1436806292672(1)$. Since the difference only occurs
in the $12^{\rm th}$ significant digit we do not feel confident that the numerical
evidence alone is sufficient to settle the matter. Ultimately we will let the reader make their own judgement.

\begin{figure}
\begin{center}
\includegraphics[scale=1.0]{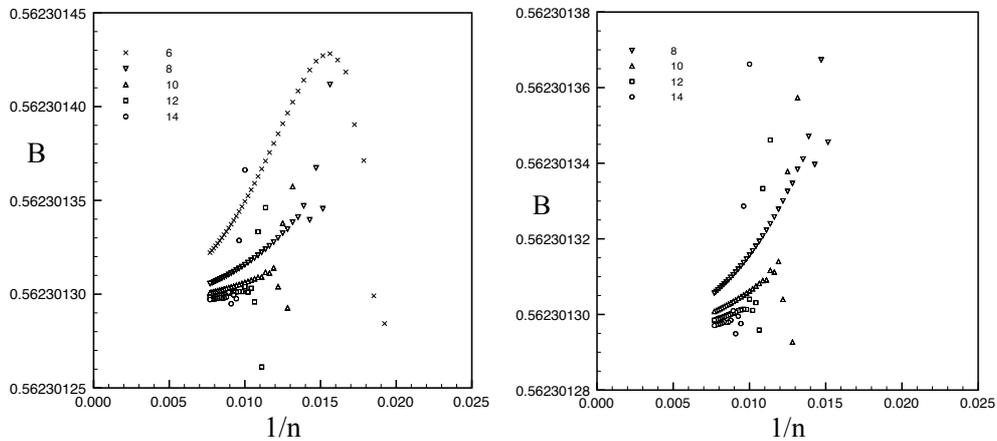}
\caption{ \label{fig:Bampl} 
Estimates for the amplitude $B$ versus $1/n$. Each data set is obtained
by fitting $p_n$ to the form (\protect{\ref{eq:sapasymp}}) using from 6
to 14 correction terms. The plot on the right is an enlarged version of
the plot to the left.}
\end{center}
\end{figure}

The detailed asymptotic form of the coefficients $p_n$ of the polygon
generating function has been studied in detail previously
\cite{Conway96,Jensen99,Jensen03}. As argued in \cite{Conway96} there
is no sign of non-analytic corrections-to-scaling exponents to the
polygon generating function and one therefore finds that
\begin{equation}\label{eq:sapasymp}
p_n = \mu^n n^{-5/2} \sum_{i\geq 0} a_i/n^i, \quad \mbox{for $n$ even}.
\end{equation}
\noindent
Estimates for the leading amplitude $B=a_0$ can thus be obtained by
fitting $p_n$ to the form (\ref{eq:sapasymp}) using increasing values
of $k$. It is useful to check the behaviour of the estimates by
plotting the results for the leading amplitude versus $1/n$ (see
Figure~\ref{fig:Bampl}), where $p_n$ is the last term used in the
fitting, and $n$ is varied from 130 down to 50. Note that as more and
more correction terms are added the estimates exhibits less curvature
and the slope become less steep. This is very strong evidence that
(\ref{eq:sapasymp}) is indeed the correct asymptotic form of $p_n$. We
estimate that $B=0.56230129(1)$.

\section{Summary and Outlook}
\label{sec:summary}
 
We have implemented a new algorithm for the enumeration of SAP on the
square lattice; the new method shows considerable promise for future enumeration studies.
The new algorithm was used to extend the series for the number of
SAP on the square lattice from $n=110$ to $n=130$.  Our analysis of the extended
series yielded improved estimates of critical parameters: $x_c^2 =
0.143680629269(2)$ ($\mu = 2.63815853035(2)$), $\alpha =
0.500000015(20)$, and $B=0.56230129(1)$.

We expect that the new algorithm will prove to be widely applicable. We chose
SAP on the square lattice for this study because it is the simplest model for computer
implementation and thus best for illustrating the basic principles involved and the
differences between  the old and new algorithms. The improvement in running time of some 15\%
while significant is unspectacular. However, we anticipate that the gains realised by the new
algorithm will be greater for other lattices, and other models such as
self-avoiding walks, theta graphs, and star polymers. It will be
especially useful in situations where many candidate completions must be
considered while pruning. 
This is certainly the case for three-dimensional lattices, for which the
restrictions on crossing of nested loops do not exist, and we are
hopeful that the new algorithm will allow for fast enumeration
of three-dimensional lattice objects via the finite lattice method. In
future, we will test this by implementing the new algorithm
for SAP and SAW on the simple cubic lattice.

 \section*{Acknowledgements}
This work was supported by an award under the Merit Allocation Scheme on the NCI National Facility at the ANU.
NC was supported by the ARC Centre of Excellence for Mathematics and Statistics of Complex Systems (MASCOS).
IJ was supported under the Australian Research Council's Discovery Projects funding scheme by the grant
DP0770705.

\end{document}